\documentclass[%
 reprint,
 amsmath,amssymb,
 aps,
 pra,
]{revtex4-2}

\usepackage{graphicx}
\usepackage{dcolumn}
\usepackage{bm}
\usepackage{lipsum}

\usepackage{booktabs}
\usepackage{multirow}
\usepackage[table,xcdraw]{xcolor}
\usepackage{array}
\begin{document}

\preprint{APS/123-QED}

\title{Optimal frequency for undulatory motion \\ in granular media}

\author{Iñaki Echeverría-Huarte} 
\author{Margarida M. Telo da Gama}%
\author{Nuno A. M. Araújo}
\affiliation{%
 Centro de Física Teórica e Computacional, Faculdade de Ciências,\\
 Universidade de Lisboa, 1749-016 Lisboa, Portugal.
}

\date{\today}

\begin{abstract}

Sand is a highly dissipative system, where the local spatial arrangements and densities depend strongly on the applied forces, resulting in fluid-like or solid-like behaviour. This makes sand swimming challenging and intriguing, raising questions about the nature of the motion and how to optimize the design of artificial swimmers able to swim in sand. Recent experiments suggest that lateral undulatory motion enables efficient locomotion, with a non-monotonic dependence of the swimming speed on the undulatory frequency and the height of the sediment bed. Here, we propose a quasi-2D granular model, where the effect of the bed height is modeled by a coarse-grained frictional force with the substrate. We show that the optimal frequency coincides with the second vibrational mode of the swimmer and explain the underlying mechanism through a characterization of the rheology of the medium. Potential implications in the design of artificial swimmers are discussed.

\end{abstract}

\maketitle


\section{Introduction}

The study of locomotion within granular media has garnered considerable scientific interest in recent decades motivated by two key factors \cite{hosoi2015beneath}. The first is the broad spectrum of natural examples. From the subterranean motion of burrowing species like earthworms \cite{dorgan2015biomechanics} to reptilian species such as sand-fish lizards \cite{maladen2009undulatory}, Nature reveals different mechanisms of efficient locomotion. These examples are sources of inspiration for the development of a new class of bio-inspired robots. The second reason is related to the non-linear properties of granular media, which exhibit a broad variety of rheological responses, resembling either solids or liquids depending on the applied forces \cite{jaeger1996physics}. Despite recent advances based on kinetic theory \cite{Brilliantov, gonzalez2024mobility}, the governing equations of granular locomotion remain elusive \cite{forterre2008flows}. 

Maladen et al. \cite{maladen2009undulatory} have shown that lizards can move under the surface of sand by performing a lateral undulatory motion, with a reported increase of the speed with the oscillation frequency. Drawing a parallelism with microorganisms swimming in fluids at low Reynolds number \cite{purcell1977life}, they adapted the principles of the classical resistive force theory (RFT) for the specific case of locomotion in granular matter \cite{zhang2014effectiveness}. They found compelling agreement between their experimental observations and the predictions of the granular RFT (gRFT), albeit with the restriction of its applicability to two-dimensional (2D) scenarios. Additional experimental limitations included the fact that lizards could not alter their swimming strategy, preventing the extrapolation of the results to other contexts.

Following this pioneering work, other examples have been published, proposing innovative experiments and potential extensions of gRFT. Peng et al. \cite{peng2016characteristics} extended the principles of gRFT to the locomotion of flexible filaments in granular media and compared them with swimming outcomes in viscous fluids. In both systems, the dynamics of the fiber provided comparable results regarding swimming efficiency, allowing the authors to extend the validated range of gRFT. Recent findings have also attempted to generalize gRFT to 3D \cite{treers2021granular}, though these approaches are known to have some limitations related with the under/over-constraints of the intruder force response \cite{agarwal2023mechanistic}. Texier et al. \cite{texier2017helical} and Valdés et al. \cite{valdes2019self} investigated the propulsion of a rotating helix in a granular medium. They found that the propulsion of the swimmer was highly influenced by the rotation speed and the shape of the helix. Specifically, both studies reported that the speed of the helix increased monotonically with the frequency, at least for the considered range of values.

Numerical models not only enable us to identify different driving mechanisms used in granular swimming but also provide a microscopic view into the grain-level dynamics. Inspired by the idea of active manipulation of the granular bed to get locomotion skills, Shimada et al. \cite{shimada2009swimming} proposed a simple model of a two-disk swimmer. By means of periodic contraction and extension processes of the two connected disks, the swimmer was able to move in a granular bed. They reported the optimal conditions to maximize swimming speed with minimal resistance. Maladen et al. \cite{maladen2011mechanical} and Ding et al. \cite{ding2012mechanics} proposed another model inspired by sandfish lizards that couples the finite element method (FEM) with discrete element modeling (DEM). FEM was employed to emulate the swimmer, while DEM represented the spherical rigid bodies constituting the granular environment. By imposing the propagation of a wave along the body of the swimmer, a strategy observed in the sandfish lizard experiments, they successfully replicated experimental results regarding swimming speed and efficiency. For the same undulatory locomotion mechanism as the two previous works, Rodella et al. \cite{rodella2020analytical} studied systematically how the oscillation parameters (i.e., amplitude and frequency) influence the swimming speed. They examined the motion of a long beam within a granular bed, discovering that within the range of oscillation frequencies studied, below the first resonance frequency of the beam, the speed of the swimmer increased monotonically.

Building on the aforementioned experimental, theoretical, and numerical studies, robots have been designed to move in granular environments. One of the first consisted of a set of self-propelled connected modules that generated an anguilliform motion within the machine, thus allowing its movement \cite{maladen2011mechanical}. Another consisted of a magnetically driven swimmer, where a magnetic head linked to a tail is able to rotate by means of an external magnetic field. As the head rotates, it causes the tail to oscillate, interacting with the granular material and propelling the robot forward. Noteworthy is the study conducted by Biswas et al. \cite{MagneticSwimmer} on a magnetoelastic robot navigating granular beds with different heights and with water as the interstitial fluid. By exploring a broad range of undulatory frequencies, they found a non-monotonic dependence of swimming speed on the oscillation frequency, which allowed the identification of the one that maximizes locomotion efficiency.

To understand this non-monotonic behavior, we introduce a quasi-2D model of granular locomotion with two basic mechanisms: (i) a bending force modulated by an oscillation resembling the self-undulatory motion of magnetic robots, and (ii) a frictional force with the substrate to account for different levels of confinement (i.e. bed heights). By conducting a systematic study across a wide range of oscillation frequencies, we recovered the non-monotonic dependence of the speed on the frequency and successfully associated the optimal frequency that maximizes speed with the second vibrational mode of the swimmer. These insights provide a validated tool and a functional initial design of efficient autonomous swimmer systems in granular environments.

The paper is organized as follows: The model is presented in Section II and the results are discussed in Section III. Finally, we draw some conclusions in Section IV.

\section{Numerical Model}

\begin{figure}[h]
\centering
  \includegraphics[width=0.45\textwidth]{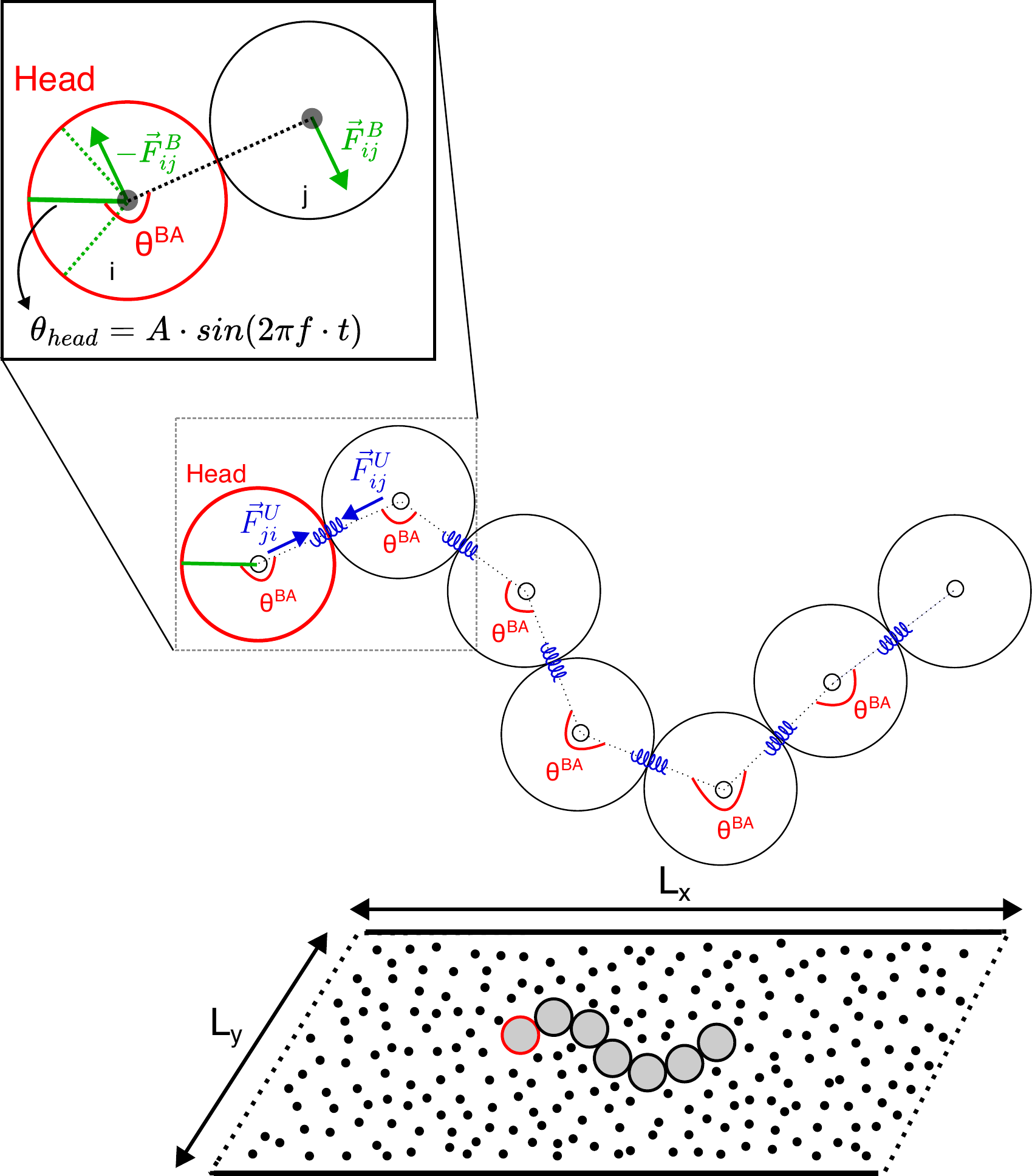}
  \caption{Schematic representation of the numerical model. A swimmer consisting of seven disks is vertically confined within a granular medium. Its constituent elements are linked together by damped springs acting in the normal direction (represented in blue). The autonomous oscillation about the axis of the head (solid green line) causes the rest of the disks to adapt to this oscillation, attempting to keep the segment that connects them (red arches) aligned. In the case depicted in the upper left corner, the two reference segments are the guiding axis of the head $\theta_{\mathrm{head}}$, and the segment that connects both particles. As mentioned, the former oscillates periodically with a maximum amplitude of $\pi/4$ (represented as dotted green lines), resulting in an angular difference $\theta^{\mathrm{BA}}$ between the two axes. The bending energy of the swimmer rectifies this misalignment through $\vec{F}_{ij}^B$, gradually making the angular difference $\theta^{\mathrm{BA}} \sim \pi$ (i.e., ensuring alignment of the segments). The same process is carried out for the remaining disks, using as reference the two axes connecting each set of trimers.}
  \label{fig:Sketch}
\end{figure}

The swimmer consists of a chain of N identical disks of radius ${r_S}$ (Fig. \ref{fig:Sketch}). The initial disk is the head (depicted in red in Fig. \ref{fig:Sketch}) while the remaining disks constitute the tail. Each one is connected to its closest neighbors leading to a damped spring force acting in the normal direction given by:

\begin{equation}
    {\vec{F}_{ij}^U} = \left[-k^U(\left|\vec{r}_{ij}\right| - 2r_S)  - \gamma^U m_{r} \cdot \vec{v}_{ij}\right] \cdot \hat{n}_{ij}
    \label{Eq:Union}
\end{equation}

\noindent where $k^U$ and $\gamma^U$ are the union spring and damping constants, respectively, $m_r = m_im_j/(m_i+m_j)$ is the reduced mass of the pairs, $m_i$ and $m_j$ the mass of each disk, $\vec{r}_{ij}$ and $\vec{v}_{ij}$ the relative position and velocity between $i$ and $j$, respectively, and $\hat{n}_{ij}$ is the normal vector connecting their centers. This approach using \emph{composite particles} has been employed in prior research \cite{Composite1,Composite2}, enabling the transfer of torques during collisions without further assumptions about surface properties such as roughness, and consequently, tangential forces.

To model the effect of the external field on the head, the axis orientation of the head of the swimmer $\theta_{\mathrm{head}}$ (illustrated by the green solid line in Fig. \ref{fig:Sketch}) is modulated using an external force. This force induces an oscillation of this axis described by the equation:

\begin{equation}
\theta_{\mathrm{head}} = A \cdot \sin(2\pi f \cdot t)
\label{Eq:ThetaEvol}
\end{equation}

\noindent where $A$ represents the amplitude, which is set to $\pi/4$ (as indicated by the green dotted lines in Fig. \ref{fig:Sketch}), and $f$ denotes the frequency of oscillation. Connecting the centers of each pair of disks defines a second axis, resulting in an angular difference $\theta_{ij}^{\mathrm{BA}}$ between them (red arches in Fig. \ref{fig:Sketch}). To ensure that each pair of axes adjusts to the oscillation of the head, we consider a pairwise bending interaction:

\begin{equation}
    \vec{F}_{ij}^{B} =  \left [-k^{B}(\pi - \theta^{\mathrm{BA}}_{ij}) - \gamma^{B} m_{r} \vec{v}_{ij} \right] \cdot \hat{t}_{ij}
    \label{Eq:Oscillation}
\end{equation}

\noindent where $k^{B}$ and $\gamma^{B}$ are the spring and damping constants, respectively, and $\hat{t}_{ij}$ is the tangential vector perpendicular to $\hat{n}_{ij}$. This force results from the bending energy of the swimmer, which tends to align all the segments connecting the disks (i.e., keep $\theta^{\mathrm{BA}}_{ij} = \pi$). Thus, referring to the example illustrated in the top-left corner of Fig. \ref{fig:Sketch}, the force ${\vec{F}_{ij}^{B}}$ acts tangentially on each pair of disks. Note that, when applying this procedure to all disks, the net force acting on the center of mass is zero and it does not move.

\subsection*{Mechanical contacts within the granular bed}

The swimmer is submerged in a granular bed, causing it to make contact with the surrounding grains as it oscillates. The swimmer-grain and grain-grain interaction forces are represented by the Hertzian spring dashpot (HSD) model \cite{HertzModel}, which characterizes the normal repulsion force between two disks upon contact as:

\begin{equation}    
    {\vec{F}_{ij}^C} = \left[-k^C \delta_{ij} - \gamma^C m_r \vec{v}_{ij}\right] \cdot \hat{n}_{ij}
    \label{Eq:Contact}
\end{equation}

\noindent where $\delta_{ij} = \left|\vec{r}_{ij}\right| - (r_1+r_2)$ is the overlap between particles or between particle and the swimmer disks, and $k^{C}$ and $\gamma^{C}$ are the spring and damping constants, respectively. 

\subsection*{Confinment}

The experimental system considered by Biswas et al. is 3D, which allows them to assess the effect of different bed heights on the dynamics \cite{MagneticSwimmer}. To mimic this effect in the 2D model,  we apply a coarse-grained frictional force of the form:

\begin{equation}
{\vec{F}_{i}^{\mathrm{conf}}} = - \mu_{\mathrm{eff}} \, W \cdot \vec{v}_i
\label{Eq:Conf}
\end{equation}

\noindent where $\mu_{\mathrm{eff}}$ is an effective friction coefficient and $W = \langle m \rangle \cdot \left| \vec{g} \right| \cdot \mathrm{(\# Layers)}$ is the magnitude of the weight bore by each particle. Here, $\langle m \rangle$ stands for the average mass of the grains, $\vec{g}$ is gravity, and $\mathrm{(\# Layers)}$ is the number of particle layers.\\

Thus, the translational motion $\vec{r}_i$ for a particle of mass $m_i$ evolves as, 

\begin{equation}
   m_i\ddot{\vec{r}}_i = {\overset{\text{Swimmer Oscillation}}{\overbrace{\sum_{j}^{\text{CP}} \left( \vec{F}_{ij}^U + \vec{F}_{ij}^{B} \right)}}} + \overset{\text{Mechanical contacts}}{\overbrace{\sum_{j}^{N_c} \vec{F}_{ij}^C}} + \overset{\text{Confinement}}{\overbrace{\vec{F}_{i}^{\mathrm{conf}}}} 
   \label{Eq:Movement}
\end{equation}

\noindent where mechanical contacts and confinement affect all particles and the first term is zero for the grains. We integrate Eq. \eqref{Eq:Movement} numerically using the Velocity-Störmer-Verlet scheme \cite{Verlet} with a time step $\Delta t = 10^{-6}\,  \mathrm{s}$. The values for the normal elastic constant $k^U$ and dissipation parameter $\gamma^U$ associated with the \emph{union} swimmer force are $k^U = 10^6\,  \mathrm{N/m}$ and $\gamma^U = 10^3\, \mathrm{s^{-1}}$. The selected values ensure that the different disks comprising the swimmer stick together across the entire range of oscillation frequencies studied. The elastic and dissipative coefficients characterizing the undulatory motion of the swimmer are specified as $k^B = 10^2\, \mathrm{N}$ and $\gamma^B =10^3 \, \mathrm{s^{-1}}$. These parameters have been calibrated to enable the observation of morphological changes in the swimmer shape at various frequencies (more details in Section \ref{Sec:Swimmer_modes}). In case of contacts, to guarantee that the overlap is less than 2\% of the particle radius, we set $k^C = 2.2 \times 10^6\, \mathrm{N/m}$ and $\gamma^C = 10^4\, \mathrm{s^{-1}}$.

\subsection*{Simulated scenarios}

To replicate the experimental setup devised by Biswas et al. \cite{MagneticSwimmer}, the swimmer comprises a total of $N = 7$ disks, each with a diameter of $d_S = 3\, \mathrm{mm}$ (resulting in a combined size of $L_s = 21\, \mathrm{mm}$) and density $\rho = 1004\, \mathrm{kg/m^3}$. We endowed the head of the swimmer with greater mass (seven times that of the other disks), ensuring that the center of mass of the swimmer is proximal to the head. The mechanical properties of the grains forming the granular bed mirror those of the swimmer, differing solely in their size $d_G$, which is uniformly distributed within the range $(1-2)\, \mathrm{mm}$.

We consider a rectangular simulation box with $L_x = 100\, \mathrm{mm}$ $(\sim 5 L_s)$ and $L_y = 50\, \mathrm{mm}$ $(\sim 2.5 L_s)$. Given the preferential motion of the swimmer along the x-axis, we implemented periodic boundary conditions along $L_x$, while confining the system vertically in $L_y$. These lateral boundaries consist of fixed particles with mechanical properties akin to the bulk grains. Ultimately, the system is randomly packed until achieving a packing fraction of $\phi = 0.8$ — slightly below the random close packing of disks in 2D (0.886) \cite{Packingfraction}. To achieve this, we initially distributed the required number of particles in a box twice the size desired. Then, employing movable walls, we compressed the granular material until reaching the targeted box size and then let it evolve to equilibrium (Fig. \ref{fig:SwimShape}a). 

Unless explicitly stated otherwise, we modified only the oscillation frequency $f$. To ensure robust statistical analysis, the results presented below are averages over at least 20 independent samples.

\section{Results}
\label{Sec:results}

\subsection{Swimming modes}
\label{Sec:Swimmer_modes}

\begin{figure}[t]
\centering
  \includegraphics[width=0.47\textwidth]{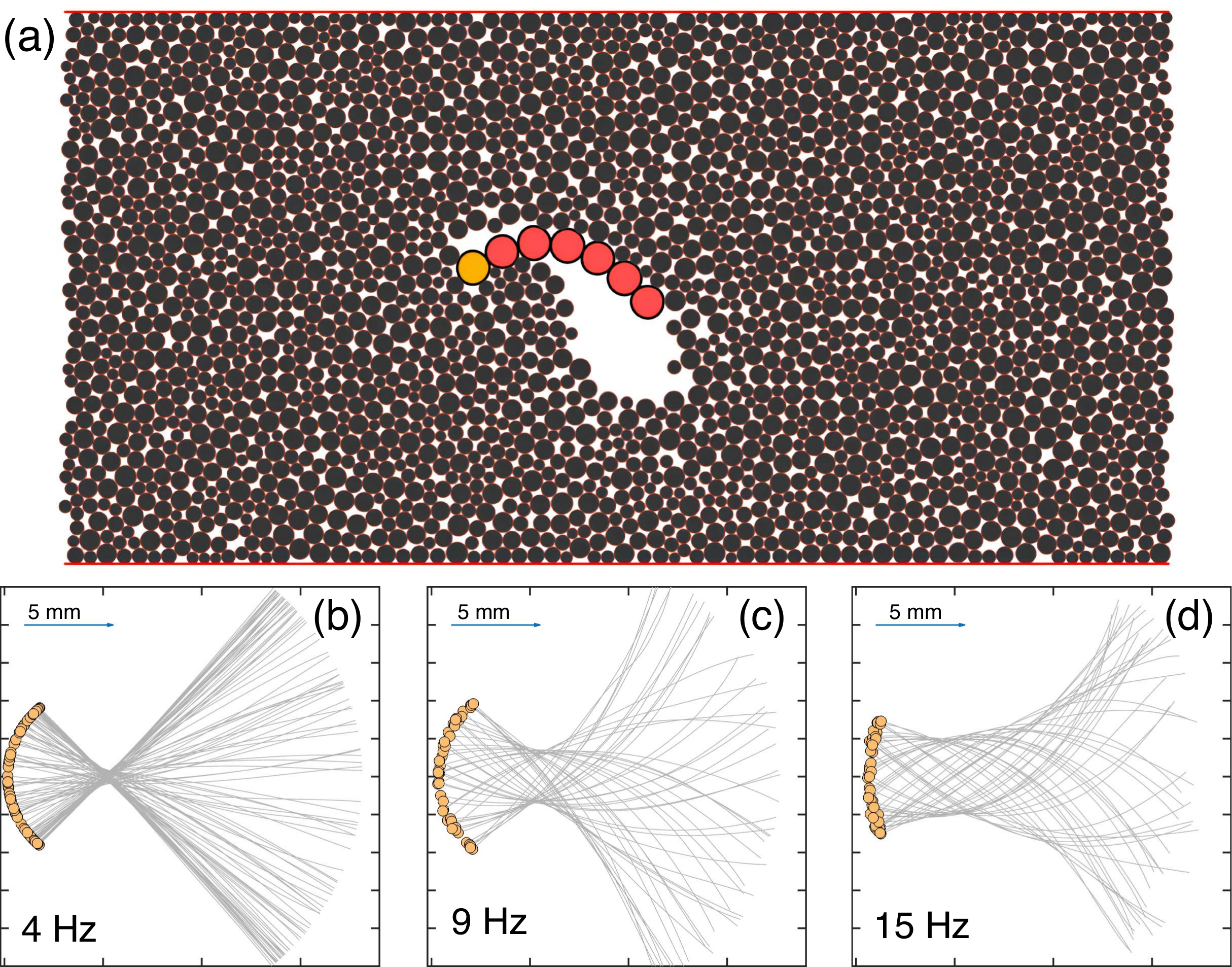}
  \caption{(a) Visualization of the simulation showcasing the motion of the swimmer oscillating at $f = 9\, \mathrm{Hz}$ within the unconfined granular medium (i.e., $\vec{F}_{i}^{\mathrm{conf}} = 0$). For reference, the head of the swimmer has been coloured in yellow. (b-d) Snapshots capturing the morphological transformations of the swimmer at various frequencies: (b) 4 Hz, (c) 9 Hz, (d) 15 Hz. All pictures are presented in the co-moving frame of the center of mass of the swimmer. These snapshots illustrate the dynamic evolution of the body shape in motion, showcasing the progression from rigid body towards anguilliform motion as the frequency increases. For clarity, the disks composing the swimmer are omitted to facilitate the overlay of shapes at different times. Instead, a line connecting the different disks is represented, with the center of the head highlighted in yellow for reference.}
  \label{fig:SwimShape}
\end{figure}

Figures \ref{fig:SwimShape}(b-d) show the shape of the swimmer for different oscillation frequencies. At low frequencies (4 Hz), since the $\theta_{\mathrm{head}}$ axis oscillates slowly, the other disks have enough time to adapt to its oscillation. Consequently, the swimmer exhibits a behavior akin to that of a rigid rod (Fig. \ref{fig:SwimShape}b, Movie S1). As the frequency increases (9 Hz), the disks connected to the head are no longer able to follow the oscillation of $\theta_{\mathrm{head}}$. A delay appears in the alignment between disks which propagates to the tail (Fig. \ref{fig:SwimShape}c, Movie S2). At high frequencies (15 Hz), the rapid orientational changes of the head axis cause a misalignment between disks, which do not have enough time to propagate to the back, resulting in a decrease in the size of the tail oscillations (Fig. \ref{fig:SwimShape}d, Movie S3). A similar behavior is observed for the experimental magnetic robot in \cite{MagneticSwimmer}.

\subsection{The effect of confinement}

\begin{figure*}[t]
\centering
  \includegraphics[width=0.78\textwidth]{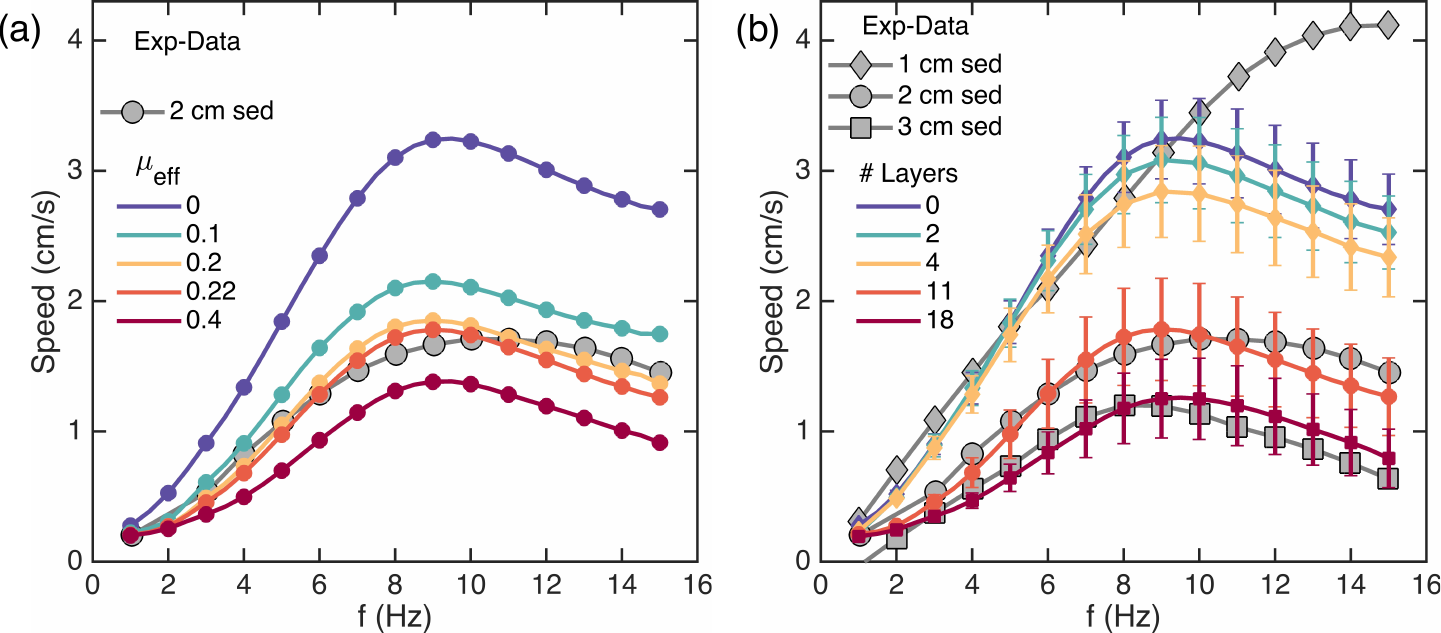}
  \caption{Estimation of $\mu_{\mathrm{eff}}$ and validation of the numerical model. (a) Swimmer speed dependence on oscillation frequency $f$. Each curve represents a different value of $\mu_{\mathrm{eff}}$ (Eq. \eqref{Eq:Conf}). The experimental data corresponds to the work of Biswas et al. \cite{MagneticSwimmer}. The objective is to find the value of $\mu_{\mathrm{eff}}$ that minimizes the mean squared difference from the experimental data. Error bars are not included for clarity. (b) Variation of the speed of the swimmer with frequency for different sediment heights. Comparison of the numerical model (colored curves) with experimental values (grey lines) \cite{MagneticSwimmer}. Error bars have been calculated as the standard error with a confidence level of 95\%.}
  \label{fig:SwimmerSpeed-ExpData}
\end{figure*}

In the experimental work of Biswas et al. \cite{MagneticSwimmer}, they configure granular beds with three different sediment heights (1, 2, and 3 cm) and measure the speed of the center of mass of the swimmer (experimental data in Fig. \ref{fig:SwimmerSpeed-ExpData}b). While for the shallowest bed, the swimming speed increases with frequency, appearing to saturate at high-frequency, in the experiments with deeper sediment beds (2-3 cm), the speed exhibits a non-monotonic behavior, revealing an optimal frequency where the propulsion is maximized.

We have examined the impact of the confinement force described in Eq. \eqref{Eq:Conf}. The aim is to determine a unique value for $\mu_{\mathrm{eff}}$ that validates the model for different number of layers and frequencies. To achieve this, we initially take as reference the curve for the 2 cm bed in Fig. \ref{fig:SwimmerSpeed-ExpData}, and estimated the number of layers—factoring in the mean dimensions of both grains and swimmer— to obtain this specific height. This gives in turn the effective weight $W$ acting on each particle. For the 2 cm bed, 11 layers closely approximate this height (3 mm for the swimmer + 11 $\times$ (1.5 mm/particle)). Once $W$ is set, we test different $\mu_{\mathrm{eff}}$ values and pick the one for which the mean squared deviation from the experimental curve is lowest. In order to conduct this comparison, we run a series of simulations wherein the oscillation frequency $f$ of the swimmer and $\mu_{\mathrm{eff}}$ are systematically varied. As depicted in Fig. \ref{fig:SwimmerSpeed-ExpData}a, it is evident that the dynamics of the system is significantly influenced by $\mu_{\mathrm{eff}}$, causing the speed of the swimmer to decrease as the effective friction increases. In the absence of this friction force ($\mu_{\mathrm{eff}} = 0$), the measured speeds are three times larger than the ones reported in the experiments. Based on these preliminary results, we estimate a value of $\mu_{\mathrm{eff}} = 0.22$. 

We can now proceed with the validation of the model for different sediment heights. In Fig. \ref{fig:SwimmerSpeed-ExpData}b, we show the performance of the model for the three sediment heights used experimentally. We found quantitative agreement between experiments and simulations for this effective friction coefficient, except for the shallowest (1 cm sediment bed), where some discrepancies are observed at high frequencies. These discrepancies are likely to result from hydrodynamic effects, neglected in the model but expected to become important for thinner sediments. Nevertheless, the model successfully captures the non-monotonic behavior of the swimming speed with frequency and the presence of an optimal oscillation that appears to be independent of the number of layers.

\subsection{Optimal Speed at Resonance Frequency}

To gain a better understanding of the non-monotonic dependence of the swimming speed with frequency, we first characterize the oscillation modes of the swimmer. We consider the unconfined system (i.e., $\vec{F}_{i}^{\mathrm{conf}} = 0$, corresponding to the case \#Layers = 0).

Let us recall that the bending mechanism is determined by the temporal evolution of $\theta_{\mathrm{head}}$ given by Eq. \eqref{Eq:ThetaEvol} and the consequent adaptation of the other disks to this oscillation through Eq. \eqref{Eq:Oscillation}. We consider a linear chain of coupled oscillators with mass $m$ where damped harmonic oscillations are driven by a sinusoidal term \cite{fowles2000analytical}, with the general form:

\begin{equation}
m \ddot{x} = - kx - c\dot{x} + f(t)
\end{equation}

\noindent where $k \equiv k^B$, $c \equiv \gamma^B\,m_r$, and $f(t)$ are determined by Eq. \eqref{Eq:ThetaEvol}. This equation is usually rewritten as:

\begin{equation}
\ddot{x} + 2\xi\omega_0 \dot{x} + \omega_0^2 x =  \frac{f(t)}{m}
\end{equation}

\noindent where $\omega_0 = \sqrt{k/m}$ is the \emph{undamped} or \emph{natural} angular frequency and $\xi$ is the damping ratio defined as $\xi = c / \left( 2 \sqrt{m\,k} \right)$. The behavior of the system critically depends on the value of $\xi$, being either overdamped ($\xi > 1$) or underdamped ($\xi < 1$). For significantly underdamped conditions ($\xi < 1/\sqrt{2}$) \cite{fowles2000analytical}, the phenomenon of resonance occurs, where for a driving frequency $\omega_r = \omega_0 \sqrt{1-2\xi^2}$, the oscillation amplitude is maximized. The swimmer is lightly damped ($\xi \approx 0.3$), and the frequency of the different resonance modes $\omega_r^n$ are readily calculated. The numerical methods used to obtain them can be found in Appendix A. 

\begin{figure}[b]
\centering
  \includegraphics[width=0.5\textwidth]{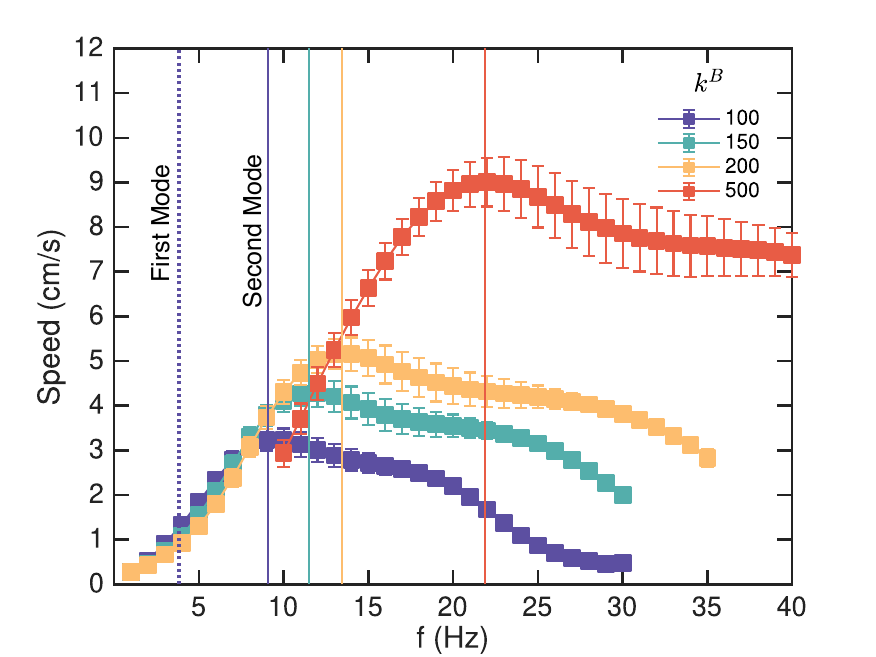}
  \caption{Swimming speed dependence on frequency for the unconfined system (i.e., $\vec{F}_{i}^{\mathrm{conf}} = 0$). Each line corresponds to a different value of $k^B$ as indicated in the legend. The vertical lines correspond to the frequencies of the resonance modes found for the swimmer characterized by $k^B$. They follow the color code in the legend. The vertical dashed line represents the first mode and the vertical solid lines the second one. For clarity, the first mode is shown only for $k^B = 10^2\, \mathrm{N}$. Error bars have been computed as the standard error at a confidence level of 95\%.}
  \label{fig:RoleofKba}
\end{figure}

\begin{figure}[h]
\centering
  \includegraphics[width=0.48\textwidth]{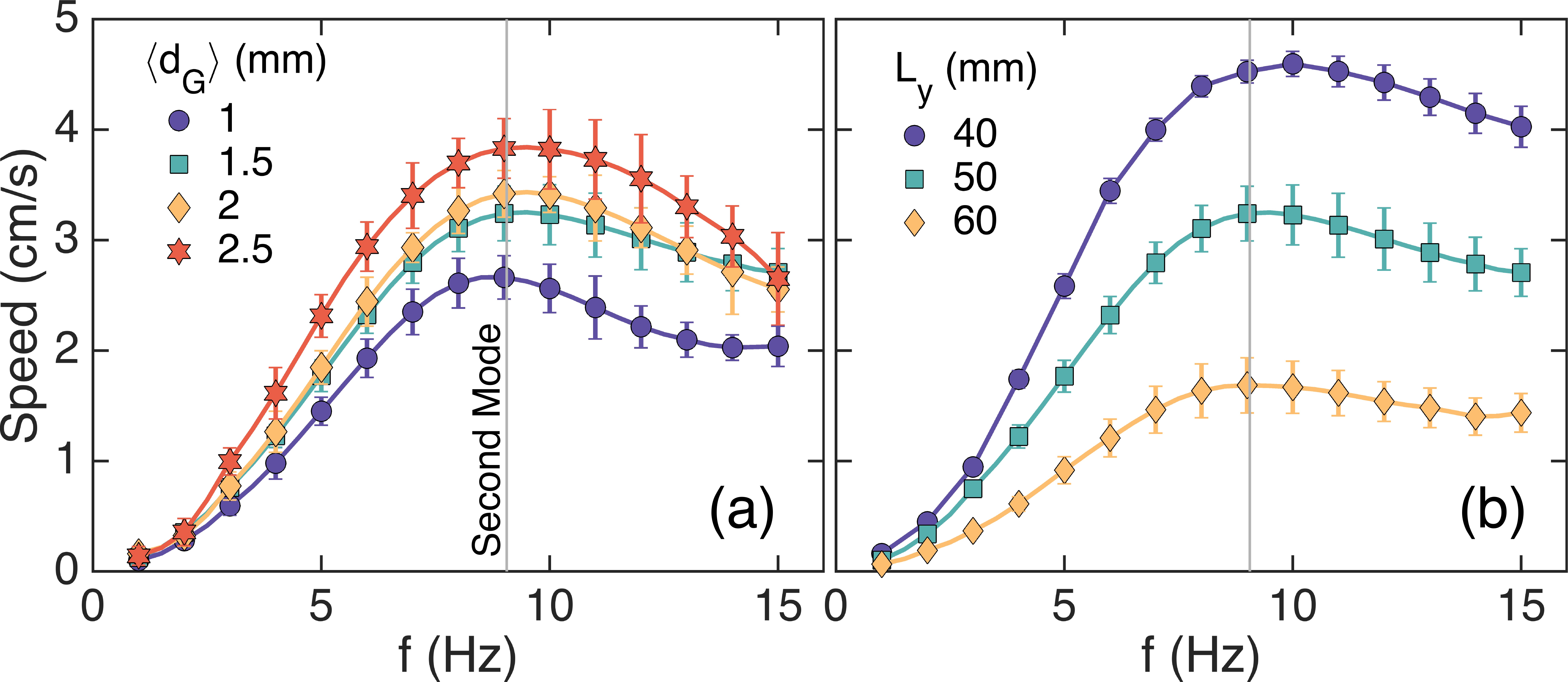}
  \caption{Swimming speed dependence on frequency for an unconfined system (i.e., $\vec{F}_{i}^{\mathrm{conf}} = 0$) changing (a) average grain size $\langle d_G \rangle$ and (b) system size $L_y$ as indicated in the legend. The vertical solid lines represent the second resonance mode. The system investigated in the main text corresponds to $\langle d_G \rangle = 1.5$ mm and $L_y = 50$ mm. The error bars have been calculated based on the standard error with a confidence level of 95\%.}
  \label{fig:EffectOf}
\end{figure}

The results in Fig. \ref{fig:RoleofKba} confirm that the frequency where the speed peaks for the swimmer with $k^B = 10^2\, \mathrm{N}$ (purple line) aligns with the second resonance mode (solid vertical line). As a further check, we ran simulations with different $k^B$. In all cases, we observed a non-monotonic behavior of the swimming speed with the oscillation frequency, with the second resonance mode aligning with the frequency that maximizes propulsion. This behavior is quite robust, regardless of whether we change the system size $L_y$ or the average grain size $\langle d_G \rangle$, as observed in Fig. \ref{fig:EffectOf}, Appendix D.

\subsection{Prevalence of the Second Resonance Mode}

\begin{figure}[t]
\centering
  \includegraphics[width=0.5\textwidth]{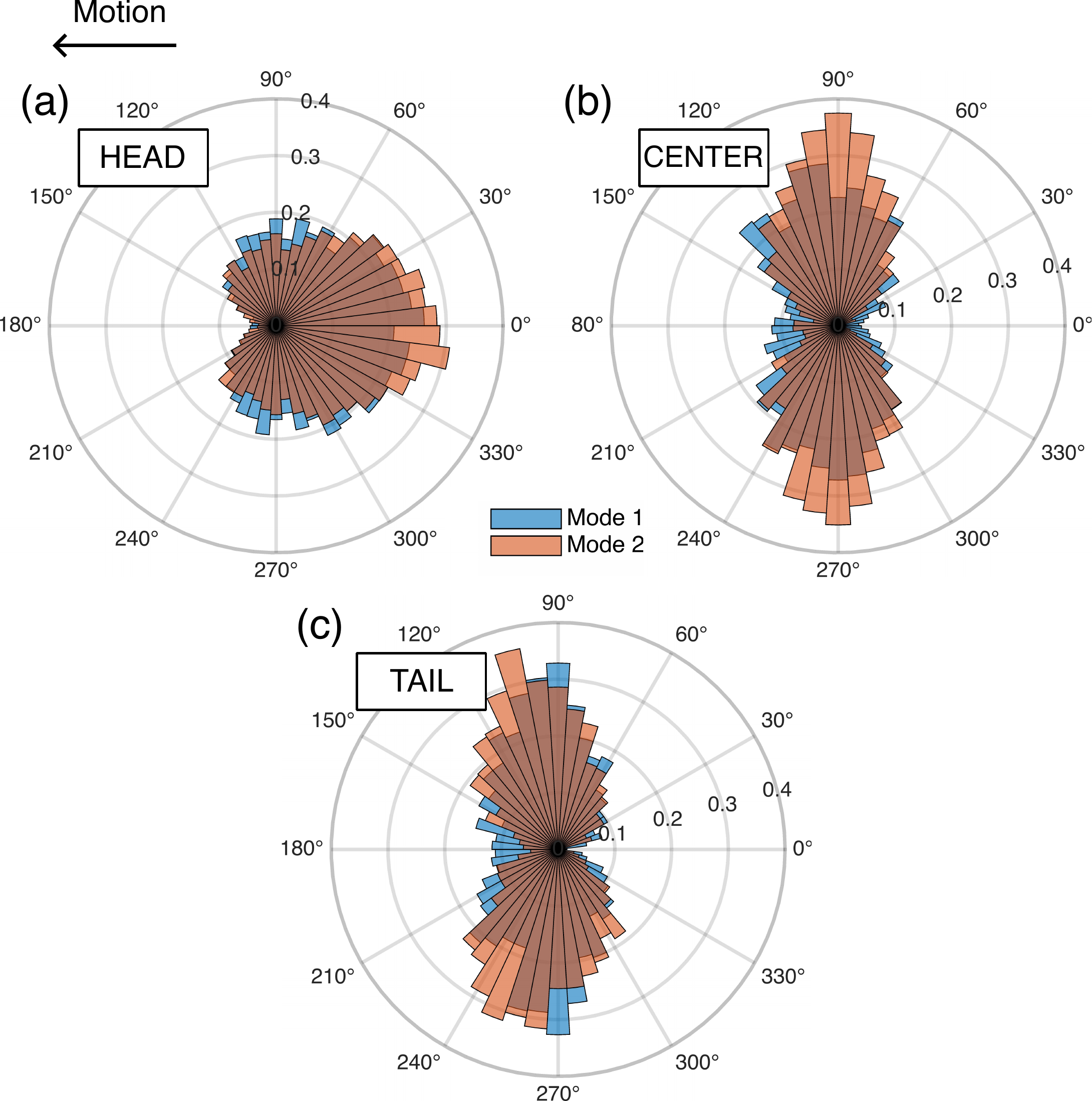}
  \caption{Polar histogram of the contact force distribution in three different regions of the swimmer: (a) head, (b) center, and (c) tail for the first two resonance modes. Blue corresponds to the first mode, orange to the second. The swimmer moves from right to left, as indicated by the arrow at the top left of the graph.}
  \label{fig:ContDist}
\end{figure}

Why is the second mode more efficient than the first one? To answer this, let us recall the changes in the shape of the swimmer with the frequency reported in Figs. \ref{fig:SwimShape}(b-c). At frequencies near the first mode ($f^1 = 3.82\, \mathrm{Hz}$), the swimmer displayed a rod-like behavior, whereas at the second ($f^2 = 9.06\, \mathrm{Hz}$), it adopted an anguilliform motion. This morphological change may have a significant effect on how the swimmer interacts with the granular bed, and thus, on the work it does to move.

In Fig. \ref{fig:ContDist} we have plotted the angular distribution of contacts between the swimmer and the grains for the two resonance modes in three regions of the swimmer; namely, the head, the center, and the tail. Starting with the head, the distributions of both modes display a similar behaviour. The contacts occur primarily against the moving direction, constituting the main region where drag forces oppose the swimmer motion. In the center, the distributions become symmetric in both modes, with most contacts pointing in the perpendicular direction to the motion. The main difference lies in the tail, where there is indeed a visible distinction between both distributions. While for the first mode contacts remain mostly vertical, in the second mode the distribution shifts in the direction of the motion. This suggests that the swimmer can gain more traction through the contacts and move faster.

\begin{figure}[t]
\centering
  \includegraphics[width=0.5\textwidth]{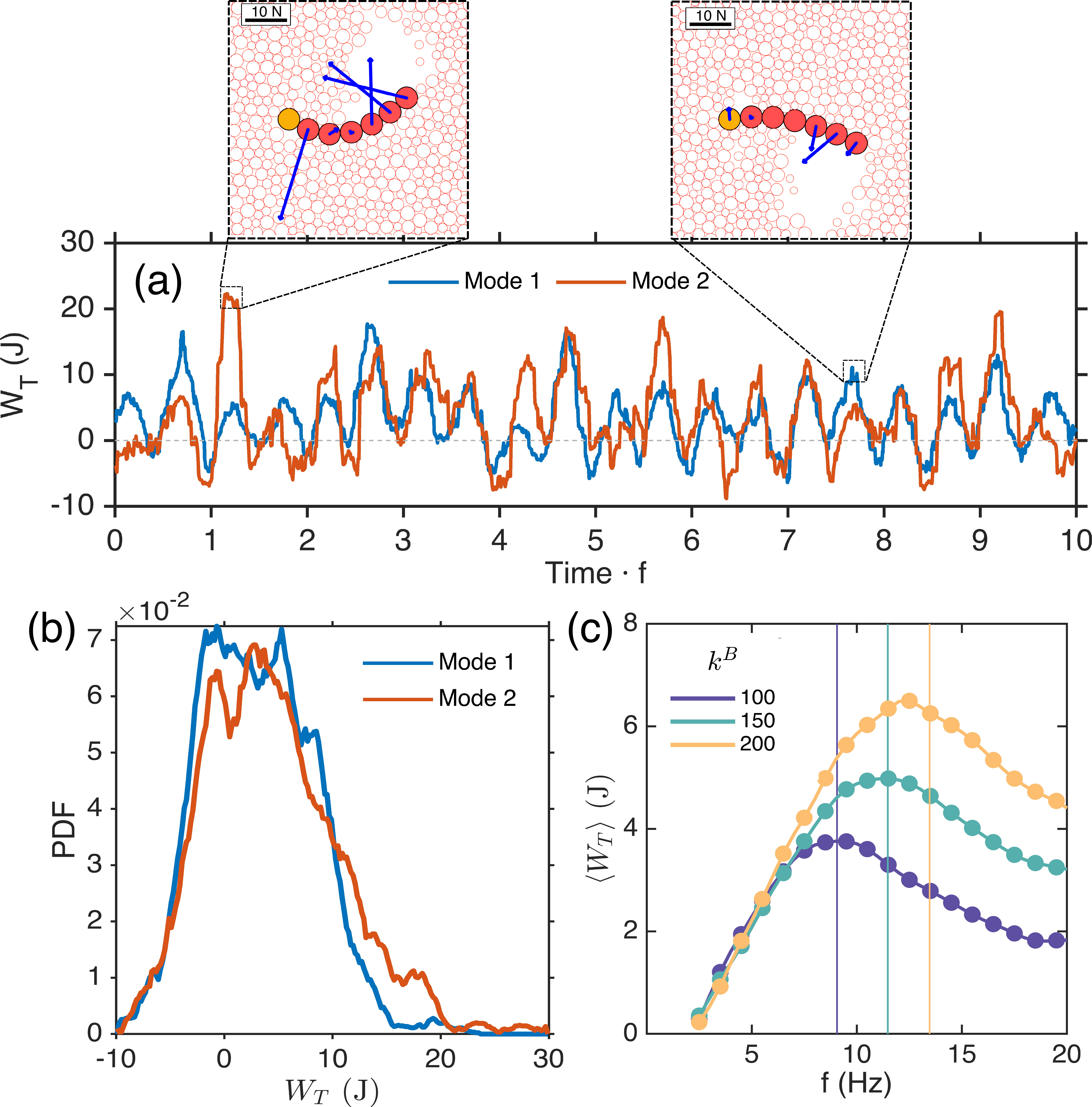}
  \caption{(a) Temporal evolution of the work done by the swimmer for the two resonance modes as indicated in the legend. To facilitate comparison, time has been rescaled by the frequency $f$, and $W_T$ is the work per cycle. The snapshots at the top illustrate the configuration at that specific moment. The blue arrows indicate the contact forces acting on the swimmer. (b) Probability density functions (PDFs) of the work $W_T$ for the two resonance modes of the swimmer. (c) Temporal average of $W_T$ for different frequencies and values of $k^B$ as indicated in the legend. The three vertical lines correspond to the second resonance mode associated with each value of $k^B$ color coded by the legend.}
  \label{fig:WorkCM}
\end{figure}

To obtain a more quantitative description, we have computed the total work done by the swimmer as:

\begin{equation}
    W_T = \int_T \vec{F}_{\mathrm{CoM}}\cdot \hat{v}_{\mathrm{CoM}}\, dt
\end{equation}

\noindent where $ \vec{F}_{\mathrm{CoM}}$ and $\hat{v}_{\mathrm{CoM}}$ are the total force and the normalized velocity of the center of mass, respectively. The integral extends over one oscillation period $T$, thus correctly scaling the work for different oscillation frequencies $f$. In Fig. \ref{fig:WorkCM}, we plot the temporal evolution of $W_T$ per cycle for the two resonance modes. In both cases, two peaks can be distinguished in each cycle where the work reaches a maximum positive value (i.e., maximum propulsion of the swimmer). These correspond to the two moments when the tail impacts the granular bed, as depicted in the illustrations on top of Fig. \ref{fig:WorkCM}. A comparison between both series suggests that the peaks of the second mode are generally higher than those of the first one. This observation can be effectively supported by examining the distribution of $W_T$ in Fig. \ref{fig:WorkCM}b, showing that for the second mode, the distribution becomes wider for $W_T > 0$. These findings allow us to confirm that, due to the morphology taken by the swimmer in the second mode, the contacts with the granular bed are more effective in propelling it than in the first mode. Finally, in Fig. \ref{fig:WorkCM}c, we compute the temporal average of the work $\langle W_T \rangle$ for different frequencies and values of $k^B$. We obtained the same non-monotonic behavior with frequency, with a peak near the peak of the swimming speed in Fig. \ref{fig:RoleofKba}.

\subsection{Granular bed behaviour}

\begin{figure}[t]
\centering
  \includegraphics[width=0.48\textwidth]{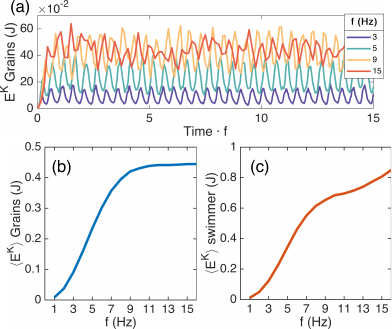}
  \caption{(a) Time series of the kinetic energy of the granular bed at different oscillation frequencies. Each line is associated with a different frequency as indicated in the legend. In (b) and (c), the dependence of the average kinetic energy on the oscillation frequency for the granular bed and the swimmer, respectively.}
  \label{fig:Energy}
\end{figure}

\begin{figure}[h]
\centering
  \includegraphics[width=0.5\textwidth]{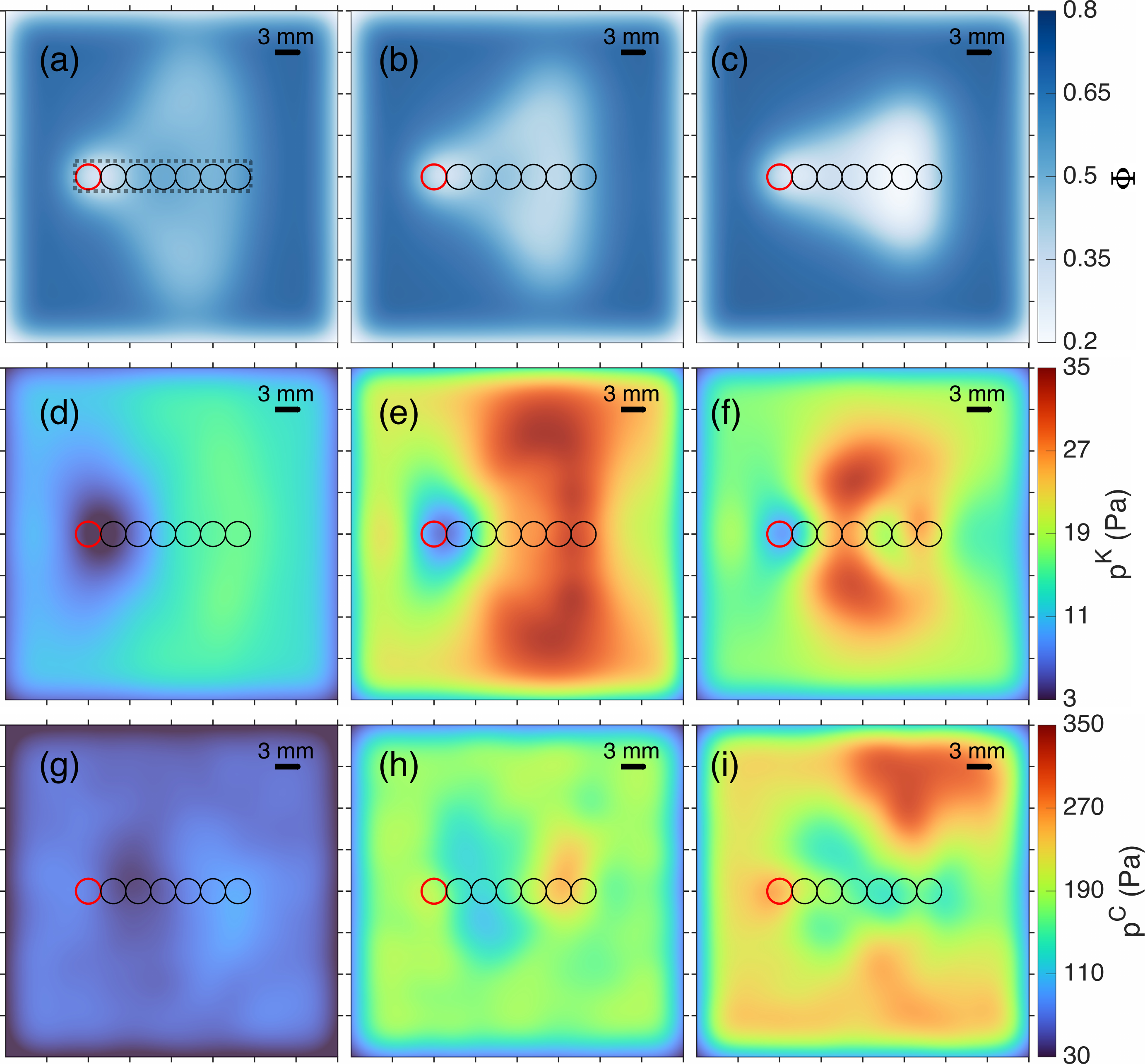}
  \caption{Solid fraction $\Phi$ (first row), kinetic pressure $p^K$ (second row) and contact pressure $p^C$ (third row) spatial profiles obtained at frequencies: (i) 4 Hz (first column), (ii) 9 Hz (second column) and (iii) 15 Hz (third column). In each row, the three plots share the same colorbar. All fields are computed in the co-moving frame of the head of the swimmer, highlighted in red for reference. The rectangle surrounding the swimmer in (a) represents the cell size used to study the spatial dependence of the fields along the vertical direction.}
  \label{fig:Fields}
\end{figure}

\begin{figure*}[t]
\centering
  \includegraphics[width=\textwidth]{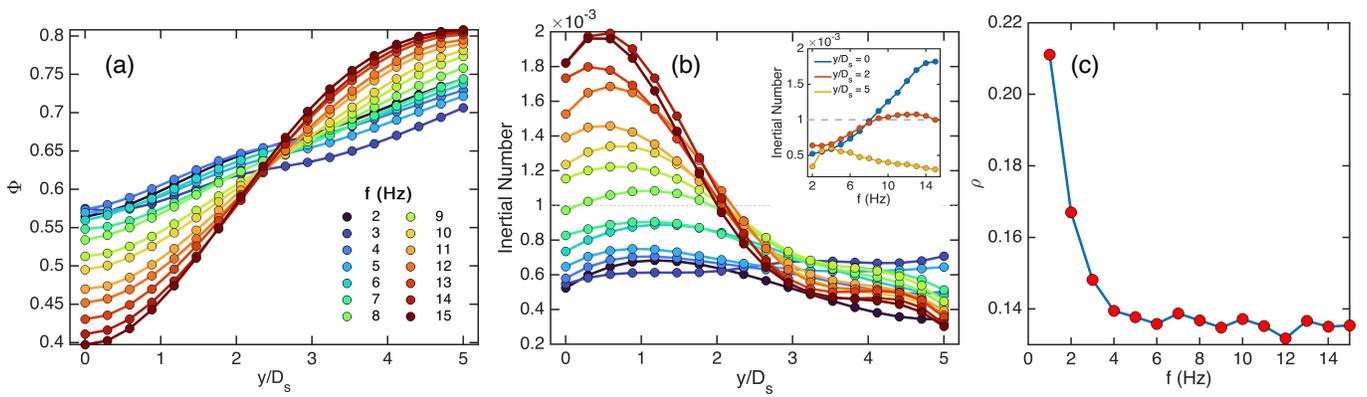}
  \caption{Vertical profiles of (a) solid fraction, and (b) inertial number (I) in a rectangular region of the size of the swimmer at different frequencies ($f$), as indicated in the legend. The vertical coordinate $y$ is normalized by the swimmer diameter $D_s$. The horizontal line at $I = 10^{-3}$ denotes the transition between the quasi-static and dense-inertial flow regimes (see main text). The inset shows the dependence of I on $f$ for the particular regions indicated in the legend. In (c), the anisotropy of the contact network ($\rho$) is plotted against the oscillation frequency.}
  \label{fig:InertialNumber}
\end{figure*}

In the granular medium, the input energy of the swimmer is dissipated due to the interaction forces resulting from collisions and relative motion between the particles. This effect can be observed in the temporal evolution of the kinetic energy of the grains in Fig. \ref{fig:Energy}a. After a short transient period that occurs when the swimmer starts oscillating, the granular bed rearranges and a stationary regime is established. Similar to the behaviour depicted in Fig. \ref{fig:WorkCM}a, the time series oscillates periodically, with two instants of energy injection per cycle corresponding to the instants when the tail of the swimmer impacts the granular bed. Regarding the maximum kinetic energy, we find that it depends on the imposed oscillation frequency. As expected, with the increase of $f$, the energy also rises. However, for $f > 9$ Hz, the differences between time series become less visible. We verified this by calculating the mean values of the kinetic energy of the granular bed. Indeed, we found that it saturates beyond a certain oscillation frequency (Fig. \ref{fig:Energy}b). This behavior is quite remarkable, considering that if we examine the kinetic energy of the swimmer, we find that it systematically increases with $f$ (Fig. \ref{fig:Energy}c). 

Aiming to further investigate the origin of the energy saturation occurring in the bulk, we conducted a rheological study within the granular medium. Through a coarse-graining mapping technique, the microscopic quantities obtained from simulations, including positions, velocities, and interaction forces, are translated into macroscopic fields (see the Appendix for further details on the coarse-graining method employed to compute these fields). Working in the co-moving frame of the head of the swimmer, the continuous fields of solid fraction $\Phi$, kinetic pressure $p^K$ and contact pressure $p^C$ are explored in a square region of size $[2L_s,\, 2L_s]$ for three different frequencies (4, 9 and 15 Hz). In all cases, the temporal average of the variables is computed along 10 s of simulation data. 

The solid fraction fields [Figs. \ref{fig:Fields}(a-c)] exhibit different spatial patterns depending on the oscillation frequency. While at $f = 4$ Hz and $f = 9$ Hz [Figs. \ref{fig:Fields}(a-b)], the region affected by the swimmer extends vertically, $f = 15$ Hz produces a more localized effect in the vicinity of the swimmer (Fig. \ref{fig:Fields}c). For the latter, the region experiences a pronounced fluidization, with $\Phi$ values significantly lower than those in the peripheral zones.

The correlation between traction transmission and velocity fluctuations can be outlined by analyzing the \emph{kinetic pressure} $p^K$. The spatial profiles in Figs. \ref{fig:Fields}(d-f) indicate that velocity fluctuations occur mainly in the tail of the swimmer and intensify with increasing frequency. However, the spatial pattern appears to be different at intermediate frequencies (Fig. \ref{fig:Fields}e) when compared to high frequencies (Fig. \ref{fig:Fields}f), akin to what has been noted in the solid fraction fields. While for the former, a profile of high fluctuations extends vertically throughout the entire region of study, the fluctuations for the latter are limited to the vicinity of the swimmer. Focusing on mechanical contacts, the spatial profiles of the \emph{contact pressure} $p^C$ are illustrated in Figs. \ref{fig:Fields}(g-i). Once again, the behavior of the bulk is determined by $f$, evolving from a homogeneous spatial distribution of contacts at $f = 4$ Hz, to $f = 15$ Hz where the contact pressure is more significant in areas far from the swimmer. 

The analysis of the fields seems to indicate that the motion of the swimmer leads to the emergence of two regions with distinct physical properties: (i) a fluidized area near the swimmer and (ii) a solid region at the borders. The characteristic sizes of these two zones will depend on $f$. For the sake of simplicity, we focus on the spatial dependence of the fields along the vertical direction within a rectangular region (of size $L_s \times D_s$) as depicted in Fig. \ref{fig:Fields}a. In Fig. \ref{fig:InertialNumber}a, $\Phi$ exhibits a crossover for different values of $f$. As the frequency increases, the transition from the liquid region (with low $\Phi$) to the solid one (with high $\Phi$) becomes more abrupt. We can quantify this effect through the so-called inertial number $I$, defined as \cite{forterre2011granular}: 

\begin{equation}
    I = \frac{\dot{\gamma}d}{\sqrt{P/\rho}}
\end{equation}

\noindent where $\dot{\gamma} = \frac{1}{2} \sqrt{(\partial_x v_y + \partial_y v_x)^2 + (\partial_x v_x - \partial_y v_y)^2}$ is the shear rate dependent on the velocity fields $v_x$ and $v_y$, $d$ is the characteristic grain size, $P$ is the pressure, and $\rho$ is the particle density. This quantity represents the ratio between a microscopic time scale associated with particle rearrangements, and the macroscopic time scale associated with the plastic deformation of the bulk. In this context, $I$ defines three important regimes: (i) quasi-static flow ($I < 10^{-3}$), which is reminiscent of a solid, (ii) dense-inertial flow ($10^{-3} < I < 1$), which resembles a fluid, and (iii) collisional (gas-like) flow ($I>10^{-1}$). In Fig. \ref{fig:InertialNumber}b, we display the spatial dependence of $I$ for different oscillation frequencies. The general trend is that $I$ decreases as we move away from the center. However, the coexistence of a liquid region with a solid one is only present for values of $f \gtrsim 8$ Hz. The interface separating these two regimes ($I = 10^{-3}$) is located approximately at a distance $y/D_s \sim 2$ from the center. We confirm that the transition between the liquid and solid regions becomes more abrupt as $f$ increases. These findings help us to conclude that the energy saturation observed in Fig. \ref{fig:Energy}b occurs due to the solidification of the medium as the oscillation frequency increases.

If we now address the dependence of $I$ on $f$ for three specific distances (inset, Fig. \ref{fig:InertialNumber}), we observe three distinct behaviors. At the center ($y/D_s = 0$), $I$ monotonically increases with frequency. By contrast, far from the swimmer ($y/D_s = 5$), the behavior is the opposite, with a monotonic decrease. At intermediate regions ($y/D_s = 2$), the behavior is non-monotonic. These different trends suggest that there must be an optimal frequency sufficient to fluidize the vicinity of the swimmer allowing its movement but also avoiding excessive solidification of the periphery, thus preventing the system from reaching a jammed state.

In recent studies on shear jamming of granular materials \cite{behringer2018physics}, it was found that the decrease in macroscopic friction is related to a reduction of the anisotropy of the contact network. We characterized the contact network through the so-called \emph{fabric tensor} (see Appendix C for details). This metric describes the microstructure of the granular bed and measures the degree of anisotropy $\rho$ in the directions of the contacts between grains. In Fig. \ref{fig:Fields}c, we plotted the dependence of $\rho$ on $f$, revealing that the anisotropy decreases until saturation as the frequency increases. This is an additional indication that, as the contact network becomes more isotropic due to the increase of the imposed pressure, the granular bed solidifies.

\section{Conclusions}

We have introduced a 2D numerical model of a swimmer capable of moving within a granular medium. Our model successfully replicates earlier experimental findings, revealing a non-monotonic behavior of the speed of the swimmer with the oscillation frequency. The validation of the numerical model allowed a systematic exploration of several parameters, some of which may be challenging to access experimentally.

The analysis of the resonance modes associated with the swimmer has confirmed that the maximum speed is attained at a frequency that corresponds to the second resonance mode. We rationalized the greater effectiveness of this mode compared to the first one by examining the contact distributions between the swimmer and the grains, as well as the work done by the former. Our analysis emphasizes how the morphological transition of the swimmer between the first and second mode results in a higher proportion of contacts aligned with the direction of motion, ultimately leading to a better propulsion efficiency. Incidentally, the prevalence of the second mode remains robust, regardless of the characteristic size of the grains or the system.

Furthermore, we have noted an intriguing energy saturation of the granular bed with increasing oscillation frequency, even though the energy of the swimmer increases monotonically. This phenomenon stems from the evolving rheological properties of the granular bed. With increasing frequency, a liquid-like region emerges near the swimmer, while a solid-like region forms farther away. As frequencies enter the intermediate range ($f \gtrsim 8$ Hz), the solid-like region becomes dominant, swiftly absorbing the energy injected by the swimmer. This absorption causes the saturation observed in the bulk. 

Here, we have neglected the effect of the interstitial fluid between the grains. This should affect the speed of the swimmer and the rheological response of the granular bed due to the presence of long-range hydrodynamic interactions that will arise in the system \cite{RodInLiquid}. However, the good agreement with the experimental results suggests that these interactions may not be crucial in highly confined scenarios. Nevertheless, this may not hold for scenarios with shallow sediment heights. Addressing this issue will be the focus of future research endeavors.

Finally, it is worth mentioning that the findings uncovered in this study may contribute to the development of new robots capable of moving in granular environments. Being able to navigate efficiently through such challenging terrains opens up possibilities for applications in fields such as search and rescue operations \cite{robot1}, environmental monitoring \cite{crover}, exploration of planetary surfaces with loose soil or sand \cite{robot2}, and even tasks in industrial settings where granular materials are involved. The insights gained from understanding the dynamics of locomotion in granular media could significantly enhance the design and performance of these robotic systems, ultimately advancing their effectiveness and versatility in real-world scenarios.


\section*{Conflicts of interest}
There are no conflicts to declare.

\section*{Acknowledgements}
The authors acknowledge financial support from the Portuguese Foundation for Science and Technology (FCT) under Contracts no. UIDB/00618/2020 (\url{https://doi.org/10.54499/UIDB/00618/2020}) and UIDP/00618/2020 (\url{https://doi.org/10.54499/UIDP/00618/2020}).

\bibliography{main}

\clearpage
\section*{Appendix}

\subsection{Normal and resonance modes of the swimmer}

\begin{figure}[h]
\centering
  \includegraphics[height=1.00cm]{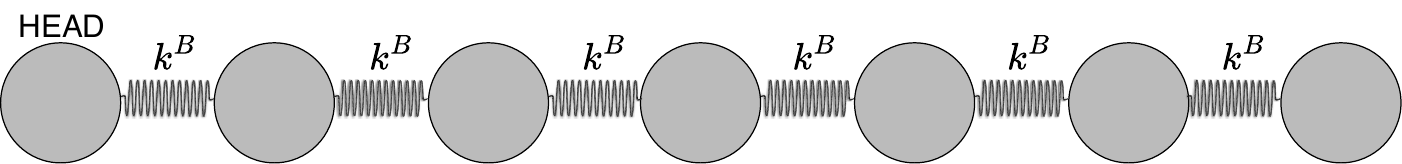}
  \caption{Representation of the linear chain of coupled oscillators.}
  \label{fig:Oscillators}
\end{figure}

To find the resonance modes of the swimmer, we use the undamped model to calculate the natural modes and then adjust the frequencies considering the damping ratio \cite{morin2002normal}. Thus, we can conceptualize the swimmer as a set of seven masses connected by six springs (Fig. \ref{fig:Oscillators}). The masses $m$ are all equal except for the mass of the head, which is $m_{h} = 7\,m$. The spring constants $k \equiv k^B$ are all equal. Let $\vec{x} = (x_1,x_2,\cdots, x_7)$ be the displacements of the masses from the equilibrium position. The equations of motion for $x_i$ are:

\begin{equation} \label{eq1}
\begin{split}
 m_{h}\, \ddot{x}_1  &= -k(x1-x2)\\
 m \,\ddot{x}_2      &= -k(x2-x1) - k (x2 - x3)\\
 m \,\ddot{x}_3      &= -k(x3-x2) - k (x3 - x4)\\
 \vdots              &  \\
 m \,\ddot{x}_7      &= -k(x7-x6)\\
\end{split}
\end{equation}

\noindent This system can be written in matrix form as:

\begin{widetext}

\begin{equation}
\renewcommand{\arraystretch}{1.0}
\begin{pmatrix}  
m_h & 0 & 0 &\cdots & 0 \\
0 & m & 0 & \cdots & 0 \\
0 & 0 & m & \dots & 0 \\
\vdots & \vdots & \vdots & \ddots & \vdots \\
0 & 0 & 0 & \cdots & m 
\end{pmatrix}
\frac{d^2}{dt^2}
\begin{pmatrix}  
x_1\\
x_2\\
\vdots\\
\vdots\\
x_7 
\end{pmatrix}
       +
\begin{pmatrix}
k&-k& 0 &\cdots &0\\
-k&2k&-k&\cdots &0\\
0&-k&2k&\cdots&0\\
\vdots&0&0&\ddots&-k\\
0&0&0&-k&k\\
\end{pmatrix}
\begin{pmatrix}  
x_1\\
x_2\\
\vdots\\
\vdots\\
x_7 
\end{pmatrix} 
=
\begin{pmatrix}  
0\\
0\\
\vdots\\
\vdots\\
0 
\end{pmatrix}
\label{Eq:MatrixForm}   
\end{equation}

\end{widetext}

or, more generally,

\begin{equation}
    M \frac{d^2 \vec{x}}{dt^2} + K\vec{x} = 0
    \label{Eq:MotionApp}
\end{equation}

\noindent where M and K are the mass and stiffness matrices, respectively. Since we are interested in finding harmonic solutions for $\vec{x}$, we can assume that the solution has the form $\vec{x} = \vec{X}\, sin(\omega t)$, and substitute it into Eq. \eqref{Eq:MotionApp}:

\begin{subequations}

    \begin{equation}
    -M \vec{X} \omega^2 sin(\omega t) + K \vec{X} sin(\omega t) = 0 \rightarrow  
    \end{equation}
    \begin{equation}
    \rightarrow K\vec{X}= \omega^2 M \vec{X}     
    \end{equation}
    
\end{subequations}

\noindent The scalar values $\lambda^n$ that satisfy a matrix equation $K\vec{u} = \lambda M \vec{u}$ are commonly known as the \emph{generalized eigenvalues} of the equation, and are directly associated with the natural frequencies through $\omega_0^n = \sqrt{\lambda^n}$. We have used MATLAB to solve the system for different values of $k^B$. The frequencies of the first and second normal modes, $\omega_0^n$, and the corresponding resonance modes, $\omega_r^n$, are listed in Table \ref{tab:ResonanceVal}. Note that to obtain the ordinary frequencies $f$, we must divide $\omega$ by $2\pi$.

\begin{table}[h]
\begin{tabular}{@{}ccccc@{}}
\toprule
$k^{B}$              & $\xi$                 & \begin{tabular}[c]{@{}c@{}}Mode \\ \textbf{n}\end{tabular} & $w_0^n$ & $w_r^n =w_0^n\sqrt{1-2\xi^2} $ \\ \midrule
\multirow{2}{*}{100} & \multirow{2}{*}{0.29} & 1                                                                           & 26.37    & 23.96                           \\
                     &                       & 2                                                                           & 62.59    & 56.91                           \\ \midrule
\multirow{2}{*}{150} & \multirow{2}{*}{0.24} & 1                                                                           & 32.30    & 30.38                           \\
                     &                       & 2                                                                           & 76.66    & 72.09                           \\ \midrule
\multirow{2}{*}{200} & \multirow{2}{*}{0.20} & 1                                                                           & 37.30    & 35.65                           \\
                     &                       & 2                                                                           & 88.52    & 84.60                           \\ \midrule
\multirow{2}{*}{500} & \multirow{2}{*}{0.13} & 1                                                                           & 58.98    & 57.95                           \\
                     &                       & 2                                                                           & 139.96   & 137.51                           \\ \bottomrule
\end{tabular}
\caption{The first and second normal modes ($\omega_0^n$) as well as the corresponding resonance modes ($\omega_r^n$) of the swimmer. Different values of $k^B$ have been used. The damping ratio ($\xi$) corresponding to each $k^B$ is provided in the second column. In all cases, the condition of significantly underdamping ($\xi < 1/\sqrt{2}$) is fulfilled, and the resonance frequency is well defined.}
\label{tab:ResonanceVal}
\end{table}

\subsection{Coarse-graining calculation}

We obtain macroscopic continuum fields by applying an integrable coarse-graining function $\varphi (\vec{r},t)$. Specifically, we have opted for a truncated two-dimensional (2D) Gaussian function $\varphi[\vec{r} - \vec{r}_i(t)] = A_{\omega}^{-1} \mathrm{exp} \left[ -(r-r_i)^2/2\omega^2\right]$, where $\omega$, the Gaussian rms width, has been selected to be equal to the the mean grain radius $\langle r_G \rangle $ of the bulk. This choice strikes a balance between being too small (which would not smooth out as desired) and being too large (which would require significant corrections \cite{CoarseGrainingTooLarge}). Here, $A_{\omega}^{-1}$ is the normalization constant ensuring that $\varphi$ is normalized in the interval $[-r_c,r_c]$, with $r_c = 3\omega$ the cut-off length. Consequently, the 2D solid fraction field $\Phi$ of the system is given at time $t$ as:

\begin{equation}
    \Phi(\vec{r},t) = A_P \sum_{i=1}^N \varphi[\vec{r} - \vec{r}_i(t)]
\end{equation}

\noindent where $\vec{r}_i$ is the position of each single grain of the granular bed and $A_p$ is the area of the grain. The velocity field $\vec{V}$ is calculated in a similar way:

\begin{equation}
    \vec{V}(\vec{r},t) = A_P \sum_{i=1}^N \vec{v}_i \, \varphi[\vec{r} - \vec{r}_i(t)] / \Phi(\vec{r},t)
\end{equation}

\noindent where $\vec{v}_i$ is the instantaneous velocity of particle i. 

Following the approach outlined in \cite{Coarse-Graining-1, Coarse-Graining-2}, we introduce the macroscopic stress field $\sigma$, which can be decomposed into a kinetic stress field ($\sigma^K$) and a contact stress field ($\sigma^C$). The kinetic stress $\sigma^K$ takes into consideration the velocity fluctuations with respect to the mean velocity field $\vec{V}(\vec{r},t)$. These fluctuations are expressed as $\vec{v}_i^{\,*}(\vec{r},t) = \vec{v}_i(\vec{r},t) - \vec{V}(\vec{r},t)$. The kinetic stress tensor is thus defined as:

\begin{equation}
    \sigma^K(\vec{r},t) = \sum_{i=1}^N \vec{v}_{i}^{\,*} \otimes \vec{v}_{i}^{\,*}\,  \varphi[\vec{r} - \vec{r}_i(t)].
\end{equation}

\noindent where $\otimes$ is the dyadic product. It is generally accepted that the trace of the kinetic stress, the so called kinetic pressure $p^K(\vec{r},t)$, is proportional to the granular temperature \cite{Irving-Kirkwood,SaraRaul}. Defined as $p^K(\vec{r},t) = \mathrm{Tr}(\sigma^K(\vec{r},t))$, it represents a measurement of the grain fluctuations respect to the mean velocity field. 

The contact stress tensor $\sigma^C$ is calculated using the interaction forces $\vec{F}_{ij}$ and branch vectors $\vec{r}_{ij}$ through the line integral of $\varphi(\vec{r},t)$ along $\vec{r}_{ij}$, as follows:

\begin{equation}
    \sigma^C(\vec{r},t) = \sum_{i=1}^N \sum_{j=1}^{N_c} \vec{r}_{ij} \otimes \vec{F}_{ij} \int_0^1 \varphi[\vec{r} - \vec{r}_i(t) + s\vec{r}_{ij}] \, ds
\end{equation}

\noindent Note that this process applies individually to every single contact $N_c$ of a particle $i$. As for the kinetic stress case, we can also define the contact pressure $p^C(\vec{r},t)$ as $p^C(\vec{r},t) = \mathrm{Tr}(\sigma^C(\vec{r},t))$ \cite{Irving-Kirkwood}. Finally, the total pressure $P$ is determined by:

\begin{equation}
    P(\vec{r},t) =  \frac{p^K(\vec{r},t) + p^C(\vec{r},t)}{2}
\end{equation}

\noindent Under stationary conditions, it is possible to characterize all these quantities by computing their mean fields. These mean fields are obtained by averaging the fields over multiple time steps in each simulation, denoted by $\Phi(\vec{r}) = \langle \Phi(\vec{r},t) \rangle$, $\vec{V}(\vec{r}) = \langle \vec{V}(\vec{r},t) \rangle$, $\sigma^K(\vec{r}) = \langle \sigma^K(\vec{r},t) \rangle$, and $\sigma^C(\vec{r}) = \langle \sigma^C(\vec{r},t) \rangle$.

\subsection{Contact Network}

The fabric tensor characterizes the contact network by analyzing the branch vectors $\vec{r}_{ij}$ \cite{Fabric1}. For systems with disk-like particles, branch vectors are defined for particle pairs in contact (i.e. when $\left| \vec{r}_{ij} \right| \leq R_{ij}$, with $r_{ij}$ denoting the inter-particle distance and $R_{ij}$ representing the sum of their radii). The fabric tensor $R$ is then defined as:

\begin{equation}
    R = \sum_{i,j}^{N_c} \vec{r}_{ij} \otimes \vec{r}_{ij}
\end{equation}

\noindent where the summation takes into account all contacts $N_c$ between pairs of particles $(i,j)$. The eigenvalues $R_1$ and $R_2$ of this tensor allow us to calculate the average number of contacts per particle $Z = R_1 + R_2$, and the anisotropy of the contact network defined as $\rho = (R_1-R_2)/Z$.

\subsection{System-size and grain-size effect}
\label{sec:EffectsOf}

\textcolor{white}{\lipsum[1]}

\end{document}